\documentclass{article}

\usepackage{amssymb}

\newtheorem{theorem}{Theorem}

\newtheorem{algorithm}[theorem]{Algorithm}
\newtheorem{corollary}{Corollary}
\newtheorem {example}{Example}
\newtheorem{definition}{Definition}

\begin{document}

\title{On an algorithm for receiving Sudoku matrices}

\author{Krasimir Yordzhev}
\date{\empty}

\maketitle

\begin{center}
{Faculty of Mathematics and Natural Sciences, South-West University\\ Ivan Mihaylov 66,
Blagoevgrad, 2700, Bulgaria\\
E-mail: {yordzhev@swu.bg}}
\end{center}

\begin{abstract}
This work examines the problem to describe an efficient algorithm for obtaining $n^2 \times n^2$ Sudoku matrices. For this purpose, we define the concepts of $n\times n$ $\Pi_n$-matrix and disjoint $\Pi_n$-matrices. The article, using the set-theoretical approach, describes an algorithm for obtaining $n^2$-tuples of $n\times n$ mutually disjoint $\Pi_n$ matrices.  We show that in input $n^2$ mutually disjoint $\Pi_n$ matrices, it is not difficult to receive a Sudoku matrix.
\end{abstract}



\section{Introduction  and notation}
Let $n$ be a positive integer. Throughout $[n]$ denotes the set $$[n] =\left\{ 1,2,\ldots ,n\right\} ,$$ $$\mathcal{U}_n =[n]\times [n] =\{ \langle a,b\rangle \; |\; a,b\in [n]\}$$ and  $\mathcal{V}_n$ denotes the set of all subsets of $\mathcal{U}_n$.

Let $P_{ij}$, $1 \leq i,j \leq n$, be $n^2$ square $n\times n$ matrices, whose entries are elements of the set $[n^2 ] =\{ 1,2,\ldots ,n^2 \}$. The $n^2 \times n^2$ matrix

$$
P =
\left[
\begin{array}{cccc}
P_{11} & P_{12} & \cdots & P_{1n} \\
P_{21} & P_{22} & \cdots & P_{2n} \\
\vdots & \vdots & \ddots & \vdots \\
P_{n1} & P_{n2} & \cdots & P_{nn}
\end{array}
\right]
$$
is called a \emph{Sudoku matrix}, if every row, every column and every submatrix $P_{ij}$, $1\le i,j\le n$ comprise a permutation of the elements of the set $[n^2 ]$, i.e., every integer $s\in \{ 1,2,\ldots ,n^2 \}$ is found just once in each row, column, and submatrix $P_{ij}$. Submatrices $P_{ij}$ are called \emph{blocks} of $P$.

This work is dedicated to the problem of finding an algorithm for getting all  $n^2 \times n^2$ Sudoku matrices for an arbitrary integer $n\ge 2$. This task is solved for $n=2$ and $n=3$ \cite{Felgenhauer}. When $n>3$, according to our information, this problem is still open. Finding algorithm to obtain $n^2 \times n^2$, $n\ge 4$ Sudoku matrices will lead to solving the problem of constructing Sudoku puzzle of higher order, which will increase the interest in this entertaining game. Here we not going to examine and compare different algorithms for solving any Sudoku puzzle. Here we will examine some algebraic properties of $n^2 \times n^2$ Sudoku matrices, which are the basis for obtaining various Sudoku puzzles.

A \emph{binary} (or   \emph{boolean}, or (0,1)-\emph{matrix})  is a matrix all of whose elements belong to the set $\mathfrak{B} =\{ 0,1 \}$.
With $\mathfrak{B}_n$ we will denote the set of all  $n \times n$  binary matrices.

Two $n\times n$ binary   matrices $A=(a_{ij} )\in \mathfrak{B}_{n}$ and $B=( b_{ij} )\in \mathfrak{B}_{n}$ will be called \emph{disjoint} if there are not integers $i,j\in [n]$ such that $a_{ij} =b_{ij} =1$, i.e. if $a_{ij} =1$ then $b_{ij} =0$ and if $b_{ij} =1$ then $a_{ij} =0$.

A matrix $A\in \mathfrak{B}_{n^2}$ is called an \emph{S-permutation} if in each row, in each column, and in each block of $A$ there is exactly one 1. Let the set of all $n^2 \times n^2$ S-permutation matrices be denoted by $\Sigma_{n^2}$.

A formula for calculating the number of all pairs of disjoint S-permutatiom matrices is given in \cite{Yordzhev20151}.

S-permutation matrices and their algebraic properties have an important part in the description of the discussed in \cite{Pallavi} algorithm.

The concept of S-permutation matrix was introduced by Geir Dahl  \cite{dahl} in relation to the popular Sudoku puzzle. It is well known that Sudoku matrices are special cases of Latin squares. It is widespread puzzle nowadays, which presents in the entertaining pages in most of the newspapers and magazines and in entertaining web sites. Sudoku, or Su Doku, is a Japanese word (or phrase) meaning something like Number Place.

Obviously a square $n^2 \times n^2$ matrix $M$ with elements of $[n^2 ] =\{ 1,2,\ldots ,n^2 \}$ is a Sudoku matrix if and only if there are  matrices $A_1 ,A_2 ,\ldots ,A_{n^2} \in\Sigma_{n^2}$, each two of them are disjoint and such that $P$ can be given in the following way:
\begin{equation}\label{disj}
M=1\cdot A_1 +2\cdot A_2 +\cdots +n^2 \cdot A_{n^2}
\end{equation}

Thus, the problem to describe an efficient algorithm for obtaining all $n^2$-tuples of mutually disjoint S-permutation matrices naturally arises. This work is devoted to this task. For this purpose, in the next section using the set-theoretical approach, we define the concepts of $\Pi_n$-matrix and disjoint $\Pi_n$-matrices. We will prove that so defined task can be reduced to the task of receiving all $n^2$-tuples of mutually disjoint $\Pi_n$-matrices.

In section \ref{sect3} we will describe an algorithm for obtaining $n^2$-tuples of $n\times n$ mutually disjoint $\Pi_n$ matrices and we will show that in input $n^2$ mutually disjoint $\Pi_n$ matrices, it is not difficult to receive a Sudoku matrix. Described in this article algorithm essentially differs from the algorithm described in [3].

\section{A representation of S-permutation matrices}\label{kikiki1}

 Let $n$ be a positive integer. If $z_1 \; z_2 \; \ldots \; z_n$ is a permutation of the elements of the set $[n] =\left\{ 1,2,\ldots ,n\right\}$ and let us shortly denote $\sigma$ this permutation. Then in this case we will denote by $\sigma (i)$ the $i$-th element of this permutation, i.e. $\sigma (i) =z_i$, $i=1,2,\ldots ,n$.

\begin{definition}\label{defPin}
Let $\Pi_n$ denotes the set of all $n\times n$ matrices, constructed such that $\pi\in\Pi_n$ if and only if the following three conditions are true:

{\bf i)} the elements of $\pi$ are ordered pairs of integers $\langle i,j\rangle$, where $1\le i,j\le n$;

{\bf ii)} if
$$\left[ \langle a_1 , b_1 \rangle \quad \langle a_2 ,b_2 \rangle \quad \cdots  \quad \langle a_n ,b_n \rangle \right]$$
is the $i$-th row of $\pi$ for any $i\in [n] =\{ 1,2,\ldots ,n\}$, then  $a_1 \; a_2 \; \ldots \; a_n$ in this order is a permutation of the elements of the set $[n]$;

{\bf iii)} if
$$\left[
\begin{array}{c}
\langle a_1 ,b_1 \rangle \\
\langle a_2 ,b_2 \rangle \\
\vdots \\
\langle a_n ,b_n \rangle \\
\end{array}
\right]
$$
is the $j$-th column of $\pi$ for any $j\in [n]$, then  $b_1 ,b_2 ,\ldots , b_n$ in this order is a permutation of the elements of the set $[n]$.

The matrices of the set $\Pi_n$ we will call \emph{$\Pi_n$-matrices}.
\end{definition}

From Definition \ref{defPin}, it follows that  we can represent each row and each column of a matrix $M\in\Pi_n$ with the help of a permutation of elements of the set $[n]$.

Conversely for every $(2n)$-tuple $$\langle \langle \rho_1 ,\rho_2 ,\ldots ,\rho_n \rangle ,\langle \sigma_1 ,\sigma_2 ,\ldots , \sigma_n \rangle \rangle,$$ where
$$\rho_i = \rho_i (1)\; \rho_i (2) \; \ldots \; \rho_i (n),\quad 1\le i\le n$$
$$\sigma_j = \sigma_j (1)\; \sigma_j (2)\; \ldots \; \sigma_j (n),\quad 1\le j\le n$$
are $2n$ permutations of elements of $[n]$ (not necessarily different), then the matrix
$$
\pi =
\left[
\begin{array}{cccc}
\langle \rho_1 (1),\sigma_1 (1)\rangle & \langle \rho_1 (2),\sigma_2 (1)\rangle & \cdots & \langle \rho_1 (n),\sigma_n (1)\rangle \\
\langle \rho_2 (1),\sigma_1 (2)\rangle & \langle \rho_2 (2),\sigma_2 (2)\rangle & \cdots & \langle \rho_2 (n),\sigma_n (2)\rangle \\
\vdots & \vdots & \ddots & \vdots \\
\langle \rho_n (1),\sigma_1 (n)\rangle  & \langle \rho_n (2),\sigma_2 (n)\rangle & \cdots & \langle \rho_n (n),\sigma_n (n)\rangle
\end{array}
\right]
$$
is matrix of $\Pi_n$. Hence
\begin{equation}\label{|Pin|}
\left| \Pi_n \right| =\left( n! \right)^{2n}
\end{equation}

\begin{definition}
We say that matrices $\pi ' =\left[ {p'}_{ij} \right]_{n\times n} \in\Pi_n$ and $\pi '' =\left[ {p''}_{ij} \right]_{n\times n} \in\Pi_n$ are \emph{disjoint}, if ${p'}_{ij} \ne {p''}_{ij}$ for every $i,j\in[n]$.
\end{definition}

\begin{definition}
Let $\pi ' ,\pi '' \in\Pi_n$, $\pi ' =\left[ {p'}_{ij} \right]_{n\times n}$, $\pi '' =\left[ {p''}_{ij} \right]_{n\times n}$ and let  the integers $i,j\in[n]$ are such that ${p'}_{ij} = {p''}_{ij}$. In this case we will say that   ${p'}_{ij}$ and ${p''}_{ij}$ are \emph{component-wise equal  elements}.
\end{definition}

Obviously two $\Pi_n$-matrices are disjoint if and only if they do not have component-wise equal elements.

\begin{example}\label{ex1}
\rm We consider the following $\Pi_3$-matrices:
\end{example}

$$
\pi' =\left[ p_{ij}' \right] =
\left[
\begin{array}{ccc}
\langle 3,1\rangle & \langle 2,1\rangle & \langle 1,2\rangle \\
\langle 2,3\rangle & \langle 3,2\rangle & \langle 1,1\rangle \\
\langle 3,2\rangle & \langle 1,3\rangle & \langle 2,3\rangle
\end{array}
\right]
$$

$$
\pi'' =\left[ p_{ij}'' \right] =
\left[
\begin{array}{ccc}
\langle 3,2\rangle & \langle 1,3\rangle & \langle 2,1\rangle \\
\langle 3,3\rangle & \langle 1,1\rangle & \langle 2,2\rangle \\
\langle 2,1\rangle & \langle 1,2\rangle & \langle 3,3\rangle
\end{array}
\right]
$$

$$
\pi''' =\left[ p_{ij}''' \right] =
\left[
\begin{array}{ccc}
\langle 3,1\rangle & \langle 1,3\rangle & \langle 2,2\rangle \\
\langle 2,2\rangle & \langle 3,1\rangle & \langle 1,1\rangle \\
\langle 2,3\rangle & \langle 1,2\rangle & \langle 3,3\rangle
\end{array}
\right]
$$

Matrices $\pi'$ and $\pi''$ are disjoint, because they do not have component-wise equal elements.

Matrices $\pi'$ and $\pi'''$ are not disjoint, because they have two component-wise equal elements: $p_{11}' =p_{11}''' =\langle 3,1\rangle$ and $p_{23}' =p_{23}''' =\langle 1,1\rangle$.

Matrices $\pi''$ and $\pi'''$ are not disjoint, because they have three component-wise equal elements: $p_{12}'' =p_{12}''' =\langle 1,3\rangle$, $p_{32}'' =p_{32}''' =\langle 1,2\rangle$, and $p_{33}' =p_{33}''' =\langle 3,3\rangle$.

The relationship between S-permutation matrices and the matrices from the set $\Pi_n$ are given by the following theorem:

\begin{theorem}\label{l2fhgg}
Let $n$ be an integer, $n\ge 2$. Then there is one to one correspondence $\theta \; :\; \Pi_n \leftrightarrows\Sigma_{n^2}$.
\end{theorem}

Proof. Let $\pi =\left[ p_{ij} \right]_{n\times n} \in \Pi_n$, where $p_{ij} =\langle a_i ,b_j \rangle $, $i,j \in [n]$, $a_i ,b_j \in [n]$. Then for every $i,j\in [n]$ we construct a binary $n\times n$ matrices $A_{ij}$ with only one 1 with coordinates $(a_i ,b_j )$. Then we obtain the matrix
\begin{equation}\label{matrA}
A =
\left[
\begin{array}{cccc}
A_{11} & A_{12} & \cdots & A_{1n} \\
A_{21} & A_{22} & \cdots & A_{2n} \\
\vdots & \vdots & \ddots & \vdots \\
A_{n1} & A_{n2} & \cdots & A_{nn}
\end{array}
\right] .
\end{equation}
According to the properties i), ii) and iii), it is obvious that the obtained matrix $A$ is $n^2 \times n^2$ S-permutation matrix.

Conversely, let $A\in \Sigma_{n^2}$. Then $A$ is in the form shown in (\ref{matrA}) and for every $i,j\in [n]$ in the block $A_{ij} $ there is only one 1 and let this 1 has coordinates $(a_i ,b_j )$. For every $i,j\in [n]$ we obtain ordered pairs of integers $\langle a_i ,b_j \rangle$ corresponding to these coordinates. As in every row and every column of $A$ there is only one 1, then the matrix $\pi =\left[ p_{ij} \right]_{n\times n}$, where $p_{ij} =\langle a_i ,b_j \rangle $, $1\le i,j\le n$, which is obtained by the ordered pairs of integers is matrix of $\Pi_n$, i.e. matrix for which the conditions i), ii) and iii) are true.
\hfill $\Box$

\begin{corollary}
Let $\pi',\pi'' \in \Pi_n$ and let $A'=\theta (\pi')$, $A''=\theta (\pi'' )$, where $\theta$ is the bijection defined in Theorem \ref{l2fhgg}. Then $A'$ and $A''$ are disjoint if and only if $\pi'$ and $\pi''$ are disjoint.
\end{corollary}

Proof. It is easy to see that with respect of the described in Theorem \ref{l2fhgg} one to one correspondence, every pair of disjoint matrices of $\Pi_n$  will correspond to a pair of disjoint matrices of $\Sigma_{n^2}$ and conversely every pair of disjoint matrices of $\Sigma_{n^2}$ will correspond to a pair of disjoint matrices of  $\Pi_n$.
\hfill $\Box$

\begin{corollary}   {\rm \cite{dahl}}
The number of all $n^2 \times n^2 $ S-permutation matrices is equal to
\begin{equation}\label{fcrl2}
\left| \Sigma_{n^2} \right| = \left( n! \right)^{2n}
\end{equation}
\end{corollary}

Proof. It follows immediately from Theorem \ref{l2fhgg} and formula (\ref{|Pin|}).
\hfill $\Box$

\section{Description of the algorithm}\label{sect3}

\begin{algorithm} \label{alg1} Receive $n^2$ mutually disjoint $\Pi_n$-matrices.

\textsc{Input:} Integer $n$

\textsc{Output:} $P_1 ,P_2 ,\ldots , P_{n^2} \in \Pi_n$ such that $P_i$ and $P_j$ are disjoint when $i\ne j$

\begin{enumerate}
  \item\label{1} Construct $n \times n$ arrays $P_1 ,P_2 ,\ldots P_n$ whose entries assume values of the set $\mathcal{V}_n$;
  \item\label{2} Initialize all entries of $P_1 , P_2 , \ldots , P_n$ with $\mathcal{U}_n$;
  \item\label{3} For every $k=1,2,\ldots ,n^2$ do loop
  \item\label{4} \hspace{0.5cm} For every $i=1,2,\ldots ,n$ do loop
  \item\label{5} \hspace{1cm} For every $j=1,2,\ldots ,n$ do loop
  \item\label{6} \hspace{1cm} Choose $\langle a,b\rangle \in P_k [i][j]$;
  \item\label{7} \hspace{1cm} $P_k [i][j]=\{ \langle a,b\rangle \}$;
  \item\label{8} \hspace{1cm} For every $t=k+1,k+2,\ldots n^2$ from the set $P_t [i][j]$ remove the element $\langle a,b\rangle$;
  \item\label{9} \hspace{1cm} For every $t=j+1,j+2,\ldots n$ from the set $P_t [i][j]$ remove all elements $\langle x,y\rangle$ such that $x=a$;
  \item\label{10} \hspace{1cm} For every $t=i+1,i+2,\ldots n$ from the set $P_t [i][j]$ remove all elements $\langle x,y\rangle$ such that $y=b$;

  \hspace{1cm} end loop \ref{5};

  \hspace{0.5cm} end loop \ref{4};

  end loop \ref{3}.
\end{enumerate}

\end{algorithm}

\begin{algorithm}\label{alg2} Receive a S-permutation matrix from a $\Pi_n$-matrix.

\textsc{Input:} $P =[\langle a_k ,b_l \rangle]_{n\times n} \in\Pi_n$, $1\le k,l\le n$.

\textsc{Output:} $S=[s_{ij} ]_{n^2 \times n^2 } \in\Sigma_{n^2}$, $1\le i,j\le n^2$.

\begin{enumerate}
  \item Construct an $n^2 \times n^2$ integer array $S=[s_{ij} ]$, $1\le i,j\le n^2$ and initialize $s_{ij}=0$ for all $i,j\in \{ 1,2,\ldots ,n^2 \}$;
  \item For every $k,l\in \{ 1, 2,\ldots , n\}$ do loop
\item \hspace{0.5cm}  $i=(k-1)*n +a_k$;
\item \hspace{0.5cm}  $j=(l-1)*n +b_l$;
\item \hspace{0.5cm}  $s_{ij} =1$

\hspace{0.5cm}  end loop.
\end{enumerate}
\end{algorithm}

\begin{algorithm}\label{alg3} Receive Sudoku matrices.

\textsc{Input:} Integer $n$.

\textsc{Output:} Sudoku matrix $A$.

\begin{enumerate}
  \item Get $n^2$ mutually disjoint $\Pi_n$ matrices $P_1 ,P_2 ,\ldots P_{n^2}$ (Algorithm \ref{alg1});
  \item For every $k=1,2,\ldots ,n^2$ from $P_k$ receive $S_k \in \Sigma_{n^2}$ (Algorithm \ref{alg2});
  \item $A=1*S_1 +2*S_2 +\cdots +n^2 *S_{n^2}$.
\end{enumerate}

\end{algorithm}

\section{Conclusion and remarks}
\begin{itemize}
  \item Described in section \ref{sect3} algorithms will work more efficiently if the programmer uses programming languages and programming environments with integrated tools for working with data structure \textbf{set} \cite{Azalov,horton2015beginning,jensenWirth,Sghildt,tan2012symbolicc++,Magda}.
  \item If in item \ref{6} of Algorithm \ref{alg1} we choose ordered pair $\langle a,b\rangle \in P_k [i][j]$ randomly, then we will get a random Sudoku matrix \cite{yordzhev_random}. Thus we tested the effectiveness of the algorithm.
  \item If in item \ref{6} of Algorithm \ref{alg1} we choose all ordered pairs $\langle a,b\rangle \in P_k [i][j]$, then finally we will get all $n^2 \times n^2$ Sudoku matrices. We do not know a general formula for finding the number $\theta_n$ of all $n^2 \times n^2$ Sudoku matrices for each integer $n\ge 2$. We consider that this is an open mathematical problem. Using a computer program based on described in section \ref{sect3} algorithms, we calculated that when $n=2$, there are $\theta_2 = 288$ number of $4\times 4$ Sudoku matrices. This number coincides with our results obtained using other methods described in \cite{yorkost}. In  \cite{Felgenhauer}, it has been shown that  there are exactly
$\theta_3 = 9! \cdot 72^2 \cdot 2^7 \cdot 27\; 704\; 267\; 971 = 6\; 670\; 903\; 752\; 021\; 072\; 936\; 960 $
number of $9\times 9$ Sudoku matrices. The next step is to calculate the number $\theta_4$ of $16\times 16$ Sudoku matrices.
\end{itemize}


\end{document}